\documentclass[10pt,conference]{IEEEtran}
\IEEEoverridecommandlockouts

\usepackage{cite}
\usepackage{amsmath,amssymb}
\usepackage{graphicx}
\usepackage{textcomp}
\usepackage{xcolor}
\usepackage{tabularx}
\usepackage{tikz}
\usetikzlibrary{arrows.meta,positioning,fit,calc}

\begin{document}

	\title{When AI Can Build It, What Should We Assess? Rethinking Evidence of Engineering Competence}
	
	\author{\IEEEauthorblockN{Ali Shakiba}
		\IEEEauthorblockA{\textit{School of Electrical and Computer Engineering} \\
			\textit{The University of Sydney}\\
			Sydney, Australia \\
			ali.shakiba@sydney.edu.au}
	}
	
	\maketitle
	
	\begin{abstract}
		Generative AI can produce code, tests, documentation, and technical explanations that resemble artefacts traditionally used as evidence of student competence, weakening the inference from artefact quality to the engineering capability behind it. We report the redesign and first deployment of a 13-week, AI-permissive, software-intensive engineering course in which students may delegate substantial work to AI while retaining responsibility for planning, verification, review, integration, and explanation. The redesign couples an \emph{engineering accountability workflow}, which structures human-AI work and preserves human decision authority, with an \emph{assessment evidence architecture} comprising six complementary channels: product, process/origin, verification, explanation, reflection, and individual assurance. A collect--cross-check--escalate (CCE) reasoning logic connects these channels rather than treating any single artefact or trace as sufficient evidence of competence. By the end of Week 11, the deployment involved 87 students organised into 17 teams, with 957 individual weekly Git repositories and 34 group-project Git repositories supporting repeated assessment and engineering activity. The first deployment yielded four practical lessons: AI assistance changes where rubric criteria carry evidential weight; multiple evidence channels require deliberate cross-checking and meaningful complementarity; the evidential strength realised in practice may differ from that intended in the assessment design; and automation scales evidence collection more readily than interpretation. The contribution is a conceptually grounded, deployed design for making engineering responsibility and competence more visible when AI can perform substantial artefact production.
	\end{abstract}
	
	
	\section{Introduction}\label{sec:introduction}
	
		Generative AI (GenAI) is challenging a long-standing assumption in engineering education: a submitted artefact provides meaningful evidence of the competencies required to produce it. Modern GenAI systems can generate, modify, test, document, and even explain software artefacts. Consequently, a technically correct program, report, test suite, or design document may no longer provide sufficient evidence, on its own, that a student can reason about, verify, integrate, or defend the engineering work embodied in that artefact.
		
		Recent studies frame this challenge as an assessment-validity issue rather than just an academic-integrity concern \cite{ebrahimzadeh2026coauthorship,bassner2026less,brattli2026validity}. Emerging evidence further suggests that AI assistance can improve task completion and the apparent quality of submissions without producing equivalent gains in conceptual understanding or independent capability \cite{bassner2026less,kuhail2024replaced}.
		
		One response is to restrict AI use or attempt to detect AI-generated work. Detection-based approaches, however, have significant limitations, particularly in cases of hybrid human-AI authorship and short-form coursework \cite{weberwulff2023testing,sun2026trusting}. 
		More fundamentally, \emph{authorship} and \emph{competence} are distinct concerns. Even a reliable estimate that AI contributed to an artefact does not establish whether a student understood, verified, adapted, or could defend the engineering decisions embodied within it. The resulting assessment challenge is therefore not only how to determine whether AI was used, but how to construct credible evidence of engineering competence when AI assistance is permitted. 
		
		This challenge is particularly consequential in software-intensive engineering education. AI-assisted development increasingly resembles a realistic condition of modern engineering practice, making it important for students to learn how to work effectively with AI-supported tools. Yet, permitting AI in open-ended engineering work complicates the attribution of competence: educators still require credible evidence that students can make engineering decisions, recognise failures, test assumptions, integrate components, and take responsibility for the resulting system. The design challenge is therefore not simply whether students should be allowed to delegate engineering work to AI, but how to ensure that delegation does not transfer responsibility for the decisions and evidence on which engineering claims depend.
		
		This paper reports the redesign and first deployment of a 13-week postgraduate engineering course around this principle. Rather than treating AI use as an exception to be prevented or detected, the course permits substantial AI-assisted work while retaining human responsibility for planning, verification, review, integration, and explanation. The redesign combines an engineering accountability workflow with an assessment evidence architecture for AI-permissive engineering work. At the end of Week 11, the deployment involved 87 students organised into 17 teams, with 957 individual weekly repositories and 34 group-assignment repositories created to support the course.
		
		The redesign builds on established principles of constructive alignment, authentic assessment, evaluative judgement, assessment validity, and the use of complementary evidence for complex competence; Section \ref{sec:related} develops these foundations and considers how GenAI changes their application. The redesign addresses three related design challenges:
		\begin{itemize}
			\item \textbf{DC1: Preserving human responsibility under delegation.} How can students delegate substantial engineering work to AI while retaining decision authority, verification responsibility, and approval over system integration?
			
			\item \textbf{DC2: Constructing credible evidence of competence.} How can assessment remain representative of engineering competence when the quality of a final artefact is no longer a sufficient proxy for the capability behind it?
			
			\item \textbf{DC3: Interpreting distributed evidence.} How can complementary but potentially inconsistent or incomplete evidence be interpreted without reducing competence to any single artefact, trace, or assessment event?
		\end{itemize}
		
		In response to these challenges, this paper makes three design contributions. First, we present an \emph{engineering accountability workflow}, a six-stage cycle (Frame $\rightarrow$ Delegate $\rightarrow$ Inspect $\rightarrow$ Verify $\rightarrow$ Review $\rightarrow$ Integrate) that structures AI delegation around explicit points of human decision authority.
		
		Second, we present an \emph{assessment evidence architecture (AEA)} comprising six complementary channels: product, process/origin, verification, explanation, reflection, and individual assurance. Each channel supports different inferences about student competence and has known limitations, allowing evidence to be interpreted in combination rather than treating any single source as sufficient.
		
		Third, we describe a \emph{collect--cross-check--escalate (CCE)} reasoning logic for interpreting evidence across these channels. Evidence is collected through normal engineering and assessment activity, cross-checked against complementary sources, and escalated for targeted follow-up when material inconsistencies or gaps remain.
		
		Beyond these design contributions, we report four lessons from the first deployment concerning the evidential value of assessment criteria, cross-channel assurance, realised evidence strength, and the human cost of interpretation. 
		
		The rest of the paper is organised as follows: Section~\ref{sec:related} reviews related work and the design foundations; Section~\ref{sec:context} describes the course context and assessment redesign; Section~\ref{sec:design} presents the coupled engineering accountability workflow, AEA, and CCE reasoning logic; Section~\ref{sec:deployment} describes their deployment and operation; Section~\ref{sec:lessons} reports lessons from the first deployment; Section~\ref{sec:transfer} discusses design implications and transferability; Section~\ref{sec:limitations} outlines the scope and limitations; and Section~\ref{sec:conclusion} concludes the paper.
	
	\section{Related Work and Design Foundations}\label{sec:related}	
		Recent work conceptualises the decoupling of artefact quality and individual capability as an assessment-validity problem arising from human-AI coauthorship. Ebrahimzadeh et al. argue that submitted work may no longer provide sufficient evidence that students understand the content they submit \cite{ebrahimzadeh2026coauthorship}. Empirical evidence from programming education reinforces this concern: Bassner et al. found that AI assistance substantially improved programming-task performance without corresponding improvements in conceptual learning or code comprehension \cite{bassner2026less}.
	
		This problem also increases the importance of \emph{evaluative judgement}: students' capability to judge the quality of their own and others' work. GenAI makes such judgement particularly important because students must evaluate not only final AI outputs but also the processes through which those outputs are produced \cite{shakiba_making_2026,bearman2024evaluative}. In software-intensive engineering, this judgement extends beyond identifying incorrect code. Students must evaluate whether generated artefacts satisfy requirements, fit system constraints, integrate with existing components, are adequately verified, and can be justified as engineering decisions. Permitting AI therefore shifts part of the educational challenge from artefact production towards maintaining meaningful human judgement, verification, and responsibility around delegated work.
	
		The need to look beyond final products is not unique to GenAI. Constructive alignment requires assessment activities to elicit evidence corresponding to intended learning outcomes rather than relying on convenient proxies \cite{biggs1996constructive}, while authentic assessment seeks alignment with the tasks, contexts, and criteria of professional practice \cite{gulikers2004authentic}. Research on professional competence similarly argues that complex capability is better inferred from a programme of complementary observations than from a single assessment instrument \cite{vandervleuten2010competence}. Evidence-Centred Design (ECD) similarly conceptualises assessment as an evidentiary argument connecting observations from assessment tasks to claims about student competence \cite{mislevy2003structure}. GenAI makes these established principles more consequential because it weakens the evidential relationship between producing an artefact and possessing the capability represented by that artefact.
	
		Software engineering education already provides several mechanisms for making development activity and individual contribution more visible. Version-control histories and project-management traces have been used to supplement conventional assessment of individual contribution in team software projects \cite{buffardi2020gitlogs}. Automated approaches can additionally combine repository activity, software behaviour, testing, and DevOps practices to provide repeated feedback and assessment evidence \cite{fernandezgauna2023devops}. However, such traces require careful interpretation: commit counts, repository activity, and automated test results reveal aspects of the development process but are not direct measures of understanding or engineering judgement. Process evidence is therefore most useful when interpreted alongside other evidence rather than treated as a replacement competence proxy.
	
		Recent GenAI-aware course designs similarly move towards assessment structures that permit AI while making human reasoning and responsibility more visible. For example, Fein et al. describe an AI-driven software-development course in which assessment considers the resulting code, the development process, and student reflection \cite{fein2026aidriven}. Related GenAI-aware assessment approaches increasingly combine AI-permissive work with mechanisms such as process-focused assessment, reflection, oral explanation, staged work, and controlled individual assessment. These designs recognise that authentic AI-assisted engineering and independent assurance serve different evidential purposes: the former can reveal how students operate in contemporary tool-rich environments, while the latter can provide stronger attribution of particular knowledge and reasoning to an individual.
		
		Contemporary responses to GenAI have introduced frameworks for both regulating AI participation and preserving assurance of individual learning. Task-level frameworks such as the AI Assessment Scale provide a structured means of specifying permissible levels of AI use, while two-lane approaches combine AI-permissive assessment with controlled conditions intended to provide stronger assurance of individual capability \cite{perkins2024aias,two_lane_2023}. However, permission and assurance structures do not by themselves specify how responsibility should be distributed \emph{within} an AI-permissive engineering task, nor how evidence generated across such work should be interpreted together.
		
		Existing work therefore establishes many of the individual elements needed for assessment under GenAI, including authentic project work, process traces, automated verification, reflection, oral explanation, and individual assurance. The gap we address concerns their system-level orchestration when substantial engineering production is explicitly delegable to AI: preserving identifiable human decision authority while constructing and interpreting complementary evidence of competence. Section \ref{sec:design} presents the resulting coupled design.
	
\section{Course Context and Redesign}\label{sec:context}
	
	The redesign was deployed in a 13-week course,  ELEC5308: Intelligent Information Engineering Practice, at the University of Sydney. The application domain is autonomous systems, but the course is software-intensive: students design modular Python systems, use version control and pull requests, build CI pipelines, test and profile software, manage experiments and data, document interfaces, and integrate perception, localisation, control, planning, and learning components. The learning outcomes emphasise analysing and evaluating methods and MLOps strategies, designing and evaluating intelligent engineering frameworks, and synthesising advanced programming pipelines. The cohort comprised 87 students organised into 17 project teams. Open-ended assignments permit AI use subject to disclosure and accountability requirements, while the final written examination is supervised and AI use is not permitted.
	
	Assignments 1 and 2 are deliberately open-ended and ill-defined, with assessment rubrics published in advance. Assignment 1 required teams to develop a modular, simulator-agnostic autonomous perception and navigation library as an installable Python package. Assignment 2 required teams to extend, refactor, or deploy that existing baseline rather than restart from scratch. This continuity was a deliberate design choice intended to approximate brownfield software development, in which earlier engineering decisions remain consequential: a well-engineered baseline becomes an asset, whereas weaknesses in architecture, testing, documentation, or integration create downstream cost. Both assignments used a Git-based collaborative workflow involving personal forks, pull requests, peer review, reproducibility requirements, and continuity between project stages.
	
	The 2026 offering involved a substantial redesign of the teaching and assessment structure. Table~\ref{tab:redesign} summarises documented structural differences between the previous offering and the 2026 design.
	
	\begin{table*}[t]
		\centering
		\caption{Assessment structure before and after the 2026 redesign.
		}
		\label{tab:redesign}
		\begin{tabular}{p{4.0cm} p{5.0cm} p{6.2cm}}
			\hline
			\textbf{Design Feature} & \textbf{Previous Offering} & \textbf{2026 Redesign} \\
			\hline
			Major assessment structure & 20\% mini-project, 40\% individual assignment/project, 40\% group project; all listed as AI-allowed & 5\% early-feedback quiz, 10\% weekly quizzes, 5\% individual weekly milestones, 20\% Assignment 1 (Group work), 30\% Assignment 2 (Group work), 30\% Supervised written final exam \\ \hline
			Supervised individual assurance & Not documented & 30\% supervised individual final examination; minimum score of 30\% on the examination required to pass the unit \\ \hline 
			Recurring individual progress gates & Not documented & Weekly individual repositories, Gradescope autograded weekly milestones, and weekly stand-up meetings (hurdle, documented but not marked) \\ \hline 
			Software engineering practices & General project/code/report/demo emphasis & Explicit MLOps, data/version control, CI/CD (encouraged where applicable), profiling, regression testing (encouraged where applicable), reproducibility, pull requests and review \\ \hline
			AI governance & AI allowed with acknowledgement following the University's general guidelines & Structured AI Journal to declare AI-use, explanation requirement, cohort-level AI-practice feedback \\ \hline 
			Project continuity & Not documented & Assignment 2 must extend/reuse/refactor the Assignment 1 baseline; fresh start is prohibited \\ \hline 
			Learning-outcome emphasis & Broad ``learn/apply'' formulation & Analyse/evaluate; design/deploy/evaluate; advanced programming pipelines \\
			\hline
		\end{tabular}
	\end{table*}
	
\section{The Coupled Design}\label{sec:design}
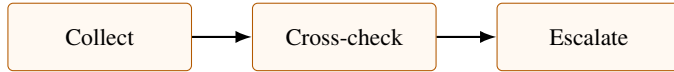
\begin{figure*}[t]
		\centering
		\resizebox{0.96\textwidth}{!}{%
			\begin{tikzpicture}[
				font=\small,
				>={Latex[length=2.2mm]},
				workflow/.style={
					draw=blue!60!black,
					fill=blue!5,
					rounded corners=2pt,
					minimum width=2.45cm,
					minimum height=9mm,
					align=center
				},
				evidence/.style={
					draw=teal!60!black,
					fill=teal!5,
					rounded corners=2pt,
					minimum width=2.45cm,
					minimum height=9mm,
					align=center
				},
				reasoning/.style={
					draw=orange!70!black,
					fill=orange!5,
					rounded corners=2pt,
					minimum width=2.45cm,
					minimum height=9mm,
					align=center
				},
				flow/.style={->,thick}
				]
				

				\node[workflow] (plan) at (0,0) {Frame};
				\node[workflow,right=0.4cm of plan] (delegate) {Delegate};
				\node[workflow,right=0.4cm of delegate] (inspect) {Inspect};
				\node[workflow,right=0.4cm of inspect] (verify) {Verify};
				\node[workflow,right=0.4cm of verify] (review) {Review};
				\node[workflow,right=0.4cm of review] (integrate) {Integrate};
				
				\foreach \a/\b in {
					plan/delegate,
					delegate/inspect,
					inspect/verify,
					verify/review,
					review/integrate}
				\draw[flow] (\a) -- (\b);
				
				\node[evidence,below=1.cm of plan] (prod) {Product};
				\node[evidence,right=0.4cm of prod] (proc) {Process / Origin};
				\node[evidence,right=0.4cm of proc] (ver) {Verification};
				\node[evidence,right=0.4cm of ver] (exp) {Explanation};
				\node[evidence,right=0.4cm of exp] (ref) {Reflection};
				\node[evidence,right=0.4cm of ref] (ass) {Individual Assurance};
				
				\node[reasoning,below left=1cm and 0.4cm of proc] (collect) {Collect};
				\node[reasoning,right=8mm of collect] (corr) {Cross-check};
				\node[reasoning,right=8mm of corr] (esc) {Escalate};
				
				\draw[flow] (collect) -- (corr);
				\draw[flow] (corr) -- (esc);
				
				\node[draw=none, align=left] at (3.75,0.95) {Engineering Accountability Workflow - How should students work with AI?};
				\node[draw=none, align=left] at (3.95,-1) {Assessment Evidence Architecture (AEA) - What evidence should be observed?};
				\node[draw=none, align=left] at (3.25,-2.85) {Assessment Reasoning (CCE) - How should evidence be interpreted?};
			\end{tikzpicture}
		}
		\caption{The coupled design. The engineering accountability workflow structures how AI-assisted work is performed and where human responsibility is retained; the assessment evidence architecture (AEA) specifies what evidence is collected; and collect--cross-check--escalate (CCE) describes how evidence is interpreted across channels.}
		\label{fig:architecture}
	\end{figure*}

	The redesign separates two related questions: \emph{How should students work with AI?} and \emph{what evidence should educators use to assess the resulting engineering competence?} The first is addressed by an engineering accountability workflow (Section \ref{sec:eng_acc_wf}), which structures delegation and preserves human responsibility at consequential decision points. The second is addressed by the AEA (Section \ref{sec:aea}), which specifies the complementary evidence used to support claims about student competence. The CCE reasoning logic (Section \ref{sec:cce}) connects these evidence channels and describes how incomplete, inconsistent, or mutually reinforcing evidence is interpreted. Figure~\ref{fig:architecture} summarises the workflow, evidence architecture, and assessment reasoning.
	
	\subsection{Engineering Accountability Workflow}\label{sec:eng_acc_wf}
		The workflow is built on a simple principle: \emph{delegate production and facilitation, not decision authority}. AI may ask questions, expose assumptions, suggest alternatives, draft plans, generate or modify code, and assist with diagnosing failures. Students, however, remain responsible for interpreting requirements, determining scope and architecture, approving decisions, inspecting generated changes, evaluating verification evidence, and authorising integration into the shared system. Table \ref{tab:authority} summarises this division of authority.
		
		The workflow establishes explicit human decision points around AI-assisted work. During planning, AI is used as a facilitator rather than as the project owner: it may surface assumptions, challenge incomplete reasoning, and suggest alternatives, but teams must review and approve the resulting decisions. During implementation, agentic tools may operate within personal development branches or forks, while changes enter the shared repository through human-reviewed pull requests. This creates a governance boundary between a relatively permissive delegation space and a human-reviewed integration space.
		\begin{table}[t]
			\centering
			\caption{Division of authority in the deployed engineering accountability workflow.}
			\label{tab:authority}
			\begin{tabular}{p{3.2cm} p{4.2cm}}
				\hline
				\textbf{AI may} & \textbf{Humans must} \\
				\hline
				Surface assumptions through structured interview & Interpret assignment requirements and constraints \\ \hline 
				Suggest alternatives and trade-offs & Select the project scope, architecture, and algorithms \\ \hline 
				Draft the master plan & Approve decisions and document unresolved disagreements \\ \hline 
				Generate or modify code & Review, inspect, and understand proposed changes \\ \hline 
				Suggest tests and potential fixes & Evaluate the adequacy of evidence and verification \\ \hline 
				Operate in a personal development fork & Review pull requests before integration into the shared system \\
				\hline
			\end{tabular}
		\end{table}

		Conceptually, the workflow is:
		\begin{equation*}
			\begin{aligned}
				\text{Frame} &\rightarrow \text{Delegate} \rightarrow \text{Inspect} \\
			             &\rightarrow \text{Verify} \rightarrow \text{Review} \rightarrow \text{Integrate}.
			\end{aligned}
		\end{equation*}
		The workflow does not prohibit substantial AI autonomy within delegated tasks. Instead, it places human decision points at boundaries where engineering responsibility must remain attributable to the student.

	\subsection{Assessment Evidence Architecture (AEA)}\label{sec:aea}
		The AEA organises assessment evidence across six complementary channels (Table \ref{tab:evidence}). These channels are distributed across the course rather than instantiated equally in every assessment activity, and no single channel is treated as sufficient evidence of engineering competence.
		
		\begin{table*}[t]
			\centering
			\caption{Evidence channels in the assessment evidence architecture (AEA).}
			\label{tab:evidence}
			\begin{tabular}{p{1.8cm} p{4.4cm} p{4.4cm} p{4.6cm}}
				\hline
				\textbf{Channel} & \textbf{Examples in the course} & \textbf{Primary inference supported} & \textbf{Known limitation} \\
				\hline
				Product & Code, models, reports, integrated system & What was produced and whether required capabilities were demonstrated & Product quality may partly reflect AI assistance rather than student capability \\ \hline 
				Process / origin & Git commits and embedded AI Journal records, forks, pull requests, issues, and planning artefacts & How the work evolved, who contributed, and where AI participation was declared & The volume of process evidence is not a reliable indicator of cognition or quality \\ \hline 
				Verification & Unit tests, CI pipelines, Gradescope autograder results, regression tests, simulation, and the 15-minute reproducibility check & What evidence supports claims that the system functions as intended & Verification activities assess only the behaviours that have been specified and tested \\ \hline 
				Explanation & Weekly stand-ups, reports, demonstrations, and short Q\&A sessions & Whether students can explain decisions, limitations, and current progress & Oral evidence is constrained by time and may vary across assessors \\ \hline 
				Reflection & Individually assessed self-reflection grounded in Git evidence and AI Journals, together with peer reflection & How students interpret their contributions, challenges, and AI-assisted decisions & Retrospective reflection can become performative or post hoc rationalisation \\ \hline 
				Individual assurance & Individual milestones, quizzes, targeted oral follow-up, and a secure final examination & What competence can reasonably be attributed to an individual under more controlled conditions & Less representative of authentic engineering practice in tool-rich environments \\ 
				\hline
			\end{tabular}
		\end{table*}

		Two distinctions are important. First, process/origin evidence is not equivalent to reflection. The AI Journal builds on a previously developed structured framework for making otherwise hidden aspects of AI-assisted engineering activity visible \cite{shakiba_making_2026}. For this course, it was further integrated into the development workflow as a Visual Studio Code extension, with journal records stored within Git commit bodies. This associates declared AI-assisted activity with a timestamped point in the repository history and the corresponding changes, allowing journal records to be interpreted alongside commits, pull requests, and other development evidence. Within the AEA, these records therefore provide process/origin evidence about where and how AI-assisted activity occurred; they are not treated as direct measures of competence or learning. By contrast, the individually assessed self-reflection asks students to interpret their contribution and learning by connecting their account to Git artefacts, task-management evidence, and specific AI Journal entries.

		Second, explanation and individual assurance support different inferences. A team demonstration may provide useful evidence that the system and its design can be explained, but it does not establish that every team member possesses the same understanding. Individual assurance therefore requires additional evidence that can be attributed to a particular student under more controlled conditions.

	\subsection{Collect--Cross-check--Escalate (CCE)}\label{sec:cce}
		The architecture becomes a coherent assessment system only when evidence from the six channels is interpreted in relation to other evidence rather than treated as a set of independent observations. We describe this assessment reasoning as CCE. CCE is not a scoring formula; it is a reasoning logic for combining complementary evidence with different strengths and limitations. CCE describes the intended reasoning structure of the assessment architecture rather than a fully standardised marking protocol. In this first deployment, evidence collection was operationalised systematically across the course, whereas cross-channel comparison and escalation were not implemented through a formal course-wide protocol with predefined triggers. Assessors nevertheless used additional inspection and targeted questioning when concerns, gaps, or inconsistencies became apparent during routine teaching and assessment activities, particularly during the weekly stand-ups. These follow-ups were recorded as part of course operations, but were initiated through assessor judgement rather than systematic CCE triage. We therefore treat CCE as the reasoning structure that emerged from and describes these practices, while recognising that its systematic operationalisation remains incomplete. This distinction between the designed architecture and its operational realisation motivates Lesson L2 (Section \ref{sec:l2}).
		
		\paragraph{Collect} Evidence is generated through both ordinary engineering activity and formal assessment, including submitted artefacts, Git and pull-request histories, test and CI records, demonstrations, reflections, and individual assessment tasks. Collection is distributed across the course so that evidence arises at multiple points in the engineering process rather than only at final submission.
		
		\paragraph{Cross-check} Evidence from one channel is interpreted in relation to other channels that support different inferences. Strong product evidence does not by itself establish strong process or collaboration evidence; a polished repository does not remove the need for reproducibility; and reflective accounts can be compared with contribution traces and declared AI use. Likewise, performance in open-ended AI-assisted project work is interpreted alongside evidence from more controlled individual assessments elsewhere in the course.
		
		\paragraph{Escalate} Material inconsistencies, gaps, or unexplained discrepancies may trigger targeted follow-up rather than being absorbed into an overall score. For example, missing or inconsistent AI-use records may lead to targeted questioning; uncertainty about individual contribution can be examined through commits, pull requests, reviews, self-reflections, and peer evidence; and submitted work that cannot be adequately explained may prompt individual oral follow-up or additional inspection.
		
		For example, consider a team submission that demonstrates strong technical functionality but is accompanied by sparse pull-request and review activity. Under CCE, the product and process/origin evidence would first be considered together rather than scored as independent observations. The assessor could then examine complementary evidence, such as verification records, timestamped AI Journal records associated with the development history, and students' explanations of the affected components. If these sources did not adequately explain the discrepancy, the case could be escalated through targeted inspection or questioning focused on the relevant engineering decisions. The purpose of escalation is not to penalise the discrepancy itself, but to obtain additional evidence where the existing channels do not support a sufficiently clear inference about the competence or contribution being assessed.

\section{Deployment and Operation}\label{sec:deployment}

	\subsection{Repeated Individual Engineering Cycles}
	
		Each week, students receive a private individual repository containing the laboratory task and supporting materials. Repository provisioning is automated through the GitHub API, substantially reducing the manual effort required to distribute and initialise repeated individual workspaces. Students complete their work within these repositories and may use AI tools where appropriate. AI Journal records were captured through the Visual Studio Code extension described in Section \ref{sec:aea} and embedded in Git commit bodies, integrating AI-use declaration into the normal repository workflow. Milestones are submitted through Gradescope, where automated checks provide rapid feedback against specified requirements such as file structure, generated outputs, unit-test behaviour, and required artefacts. The repositories also preserve each student's development history, providing repeated evidence of process, verification, and individual progress while giving students regular practice with software-engineering and version-control workflows.
	
		By the end of Week 11, 957 individual weekly repositories had been created, in addition to 34 group-project repositories supporting Assignments 1 and 2. From Week 4 onwards, the teaching team also conducted weekly stand-up-style discussions with each project team. Staff rotated between groups and asked each student what they had completed, what they were currently working on, and whether they were encountering blockers. These discussions typically lasted around 10 minutes per team but could extend beyond 20 minutes when substantial questions or technical difficulties arose. The stand-ups were formative rather than separately marked; their primary purposes were to provide recurring oral visibility of individual progress, surface difficulties early, and create opportunities for timely intervention. Notes from these discussions were retained as supporting evidence when later addressing concerns about group dynamics or contribution.
		
	\subsection{Project Workflow and Assessment Continuity}
		\emph{Assignment 1} is the first major integration point. Teams develop a modular autonomous perception and navigation library designed to serve as the baseline for Assignment 2. The specification requires an installable Python package, a clean API, a simulator-independent core, unit testing, documentation, installation guidance, a worked example, and Git-based collaboration. Reproducibility is operationalised explicitly: a new user should be able to clone the repository and run the submitted demonstration on a supported machine within 15 minutes.
		
		The marking process enforces this threshold. Assessors use a shared rubric discussed in a calibration meeting and attempt to reproduce and inspect the submitted system. Failure to achieve successful setup within 15 minutes is penalised in the relevant criteria, although assessors may continue troubleshooting to inspect remaining functionality.
		
		Process evidence is explicitly assessed rather than collected only for visibility. Collaboration and repository criteria include meaningful commits, pull requests and reviews, evidence of shared contribution, and professional repository organisation. Students work through personal forks and focused branches, with peer review expected before changes are merged into the group repository or the main branch. Weak process evidence can therefore reduce marks even when technical functionality is strong.
		
		The assignment concludes with an approximately eight-minute demonstration followed by a two-minute Q\&A. Although the specification states that every team member should be able to answer questions about any component, the available Q\&A time usually allowed only one, and occasionally two, students to respond. We therefore interpret this activity as providing limited team-level explanation evidence rather than systematic individual assurance.
		
		Students also submit an individually marked self-reflection covering contribution evidence, a learning or challenge episode, and AI use grounded in specific journal entries. In addition, students provide structured feedback to at least two other teams on strengths, potential improvements, useful questions for further examination, and ideas transferable to their own project. Each team receives this aggregated feedback anonymously as GitHub issues carried forward into Assignment 2.
		
		\emph{Assignment 2} deliberately begins from the Assignment 1 baseline rather than allowing a clean restart. Teams must reuse, refactor, or extend the existing library into a more complete autonomous system. The task remains intentionally open-ended: teams select their target capability and may work in different simulation or hardware environments, but must justify scope, design choices, evidence, and limitations.
		
		Before development, teams submit a short proposal describing the intended deliverable, platform, reuse of the Assignment 1 baseline, technical risks, and expected demonstration evidence. Scope feedback is returned before substantial implementation begins, allowing feasibility and risk to be addressed prospectively. Assignment 1 feedback is also embedded into the development environment through GitHub issues where appropriate, so identified weaknesses become actionable engineering tasks rather than post-hoc comments. The two assignments therefore create a continuity chain from baseline architecture, through assessment feedback, to open-ended extension.
	
		\subsection{AI-Practice Feedback and Agentic-Development Instruction}
		
			The AI Journals were used not only to document AI-assisted activity, but also as a source of aggregate evidence about emerging AI-assisted engineering practices. Building on their combination of interaction tracking and reflective auditing~\cite{shakiba_making_2026}, the teaching team used recurring cohort-level patterns to inform feedback sessions in Weeks 4 and 8.
			
			The Week 4 intervention focused on prompting: providing task context, concrete evidence such as tracebacks or outputs, explicit constraints, expected behaviour, and verification criteria rather than relying on broad or context-free requests. By Week 8, the emphasis had shifted from prompt construction toward agentic engineering, including providing agents with relevant artefacts rather than only textual descriptions, reviewing generated changes, testing modifications through available verification mechanisms, and making explicit adopt, override, or could-not-verify decisions.
			
			These interventions were aimed to form a course-level formative loop:
			\begin{equation*}
			\begin{aligned}
			\text{Observe} &\rightarrow \text{Aggregate} \rightarrow \text{Teach Back} \\
	              &\rightarrow \text{Raise Expectations} \rightarrow \text{Observe Again}.
			\end{aligned}
			\end{equation*}
			The second half of the course extended this loop from formative feedback into explicit instruction on agentic software development. Part of the Week 9 lecture addressed task and execution boundaries, code ownership and review responsibilities, repository-level agent instructions and skill files, human approval before integration, and systematic inspection of agent-generated changes.
			
			A third cohort-level AI-practice reflection is scheduled for Week 13 to revisit how students' approaches to delegation, verification, review, and agent governance have evolved across the semester. Because this activity had not yet occurred at the reporting point, it is treated as planned course design rather than deployment evidence.
	
	\subsection{Individual Assurance and Escalation}
		The redesign does not attempt to make every assessment activity AI-resistant. Open practical work and team projects assess performance in a modern tool-rich environment, while individual quizzes, weekly milestones, targeted oral follow-up, and the supervised final examination provide evidence that can be attributed more directly to individual students. The final examination accounts for 30\% of the unit mark and includes a hurdle requirement: students must achieve at least 30\% of the available examination marks to pass the unit.
	
		These components support two complementary claims: (1) whether students can engineer effectively when AI tools and collaboration are available, and (2) what knowledge and reasoning can reasonably be attributed to an individual under more controlled conditions. The architecture treats these claims as complementary rather than substitutable.
		
		Where evidence from open project work is incomplete or inconsistent, targeted inspection or oral follow-up can be used to obtain additional individual evidence. Escalation is therefore selective rather than universal: controlled assessment provides a baseline of individual assurance, while further follow-up is reserved for cases in which other evidence remains insufficient or contradictory.
		
	
	\subsection{Operational Scale and Workload}
		Automation made repeated repository provisioning, evidence collection, and autograding operationally feasible, but it did not make the redesign inexpensive to run. Each tutor was allocated two two-hour practical sessions, two hours of preparation, and a one-hour coordination meeting per week. The coordinator's workload was not formally time-tracked, but the first deployment required substantial additional effort for lecture redevelopment, laboratory and assessment design, repository automation, autograder development, AI-Journal tooling and processing, coordination, online support, and project feedback. Much of this work was concentrated in a single coordinating role. At this scale, automation supported the management of hundreds of repositories and repeated evidence-collection activities across the 87-student cohort.
		
		Assessors were also provided with automated repository-analysis tools developed by the course coordinator that summarised commit histories, pull requests, reviews, contribution distributions, document lengths, review behaviour, merge patterns, and selected software-quality indicators used in the assessment rubrics. These tools reduced the effort required to locate potentially relevant evidence, but they did not replace assessor judgement about its meaning or adequacy.

\section{Lessons Learned}\label{sec:lessons}

	\subsection*{L1: AI Changes Where Assessment Criteria Carry Evidential Weight}\label{sec:l1}
		A common assumption in AI-permissive course design is that AI shifts the difficulty of assessment from artefact production towards higher-order reasoning. However, the first deployment suggests a more specific assessment-design challenge: \emph{AI assistance does not affect all rubric criteria equally}. Criteria that traditionally function as quality floors, such as whether software can be installed, whether a repository is well organised, or whether code exhibits basic structural separation, became comparatively easier for students to satisfy with AI assistance. By contrast, criteria involving system integration, cross-component behaviour, failure diagnosis, and the justification of engineering trade-offs continued to reveal substantial differences in the quality of students' work.
		
		This matters because rubric weights established for pre-GenAI conditions may no longer reflect where meaningful evidence of competence is most visible. If substantial weight remains attached to criteria that AI can readily help students satisfy, while comparatively little weight is placed on integration, verification, diagnosis, and engineering decision-making, higher artefact quality may contribute disproportionately to the final mark without providing correspondingly strong evidence of student capability.
		
		The Assignment 2 rubric was revised in response to this observation. Integration quality and engineering improvement over the Assignment 1 baseline were given greater prominence, and students were required to provide explicit evidence of integration-stage decisions, including where AI-generated components required adaptation, what was verified before integration, and what was rejected or overridden. The resulting lesson is not that AI-assisted criteria should be removed, but that rubric weight should be reconsidered according to where human engineering judgement remains visible and consequential.
		
	\subsection*{L2: Multiple Evidence Channels Do Not Automatically Create Cross-Checking}\label{sec:l2}
		The deployment revealed two related limitations of a multi-channel architecture. First, making several evidence channels available does not ensure that they will be systematically cross-checked. Second, cross-checking is only informative to the extent that the channels provide meaningfully complementary evidence.
		
		The CCE logic assumes that evidence from different channels is interpreted together rather than treated as a set of parallel observations. The first deployment exposed a gap between this design intent and operational practice. Cross-channel follow-up did occur: during weekly stand-ups, assessors sometimes examined additional evidence or questioned students when concerns or inconsistencies became apparent. However, these cases arose through assessor judgement rather than a systematic course-wide triage procedure. At a cohort size of 87 students, and under a substantial first-deployment workload, evidence collection was therefore much easier to operationalise than systematic cross-channel interpretation. AI Journals, oral Q\&As, repository histories, reflections, and product evidence were all available for such follow-up, but relating them systematically across the cohort remained difficult.
		
		One consequence was that potentially meaningful configurations of evidence were difficult to identify systematically. For example, no single channel may indicate a problem, yet a combination of unusually polished product evidence, sparse development traces, highly generic AI-use records, and weak explanation may warrant closer inspection. Detecting such patterns requires assessors to compare channels explicitly rather than interpret each one independently. In a future deployment, this comparison should be supported through pre-processing or structured triage so that cross-checking becomes a distinct phase of assessment rather than an implicit expectation placed on individual assessors. Candidate triage rules include flagging strong product evidence accompanied by sparse development or review traces; reflection or AI-use accounts that are difficult to reconcile with observable repository activity; and claims of verification that are not supported by corresponding test, CI, or reproducibility evidence. Such indicators should trigger closer cross-channel inspection rather than directly determine marks, preserving assessor judgement while concentrating limited attention on cases where complementary evidence is most informative.
		
		A second lesson concerns the independence of evidence channels. Six channels do not necessarily provide six independent sources of assurance. A student may use AI to assist with code, journal entries, reflective writing, and prepared explanations, making these channels partly correlated outputs of the same AI-mediated process. Evidence is therefore stronger when channels depend on meaningfully different capabilities or operating conditions. 
		
		The practical implication is that evidence architecture should be designed not only around the number of channels, but around their complementarity and independence. Replication therefore requires explicit resources for cross-channel comparison and deliberate inclusion of evidence sources that cannot all be satisfied through the same AI-assisted workflow. In the next offering, we plan to allocate dedicated time within tutors' weekly workload to review each team's repository activity, including commits, pull requests, and review histories, before stand-up meetings. We also plan to provide specialised repository-analysis tools that summarise potentially relevant evidence and help tutors identify issues requiring discussion. In the first deployment, repository activity was reviewed weekly by the course coordinator, leading to targeted intervention in three instances; however, the available tutor workload did not support comparable preparation and systematic discussion across all teams. The planned change therefore shifts cross-channel inspection from a coordinator-centred activity towards a resourced part of tutors' preparation for evidence-informed stand-up meetings.
		
	\subsection*{L3: Designed Evidence Strength May Differ from Realised Evidence Strength}\label{sec:l3}
		The first deployment showed that the intended evidential role of an assessment activity may differ substantially from the evidence it produces under real operating constraints. The Assignment 1 demonstration was designed to provide explanation evidence and to contribute to individual assurance: all team members were expected to understand the complete system and to be able to answer questions about any component. In practice, however, the two-minute Q\&A window usually allowed only one, and occasionally two, students to respond.
		
		The activity therefore provided useful team-level explanation evidence but much weaker individual evidence than intended. This limitation arose from an operational constraint: given the number of teams scheduled within each practical session, substantially more than approximately ten minutes per team could not realistically be allocated to the combined demonstration and Q\&A. The available time therefore limited the extent to which individual understanding could be sampled across all team members.
		
		This constraint could not be removed within the current offering. As a partial mitigation, the Week 12 practical is repurposed as a formative help-desk session. Students first will complete a 40-minute mock final examination as their individual weekly quiz and receive feedback in preparation for the supervised final exam. During the remaining practical time, each team works with its assigned assessor to walk through its codebase, deliverables, and system components, discuss the rubric, and identify missing or incomplete elements. No marks will be assigned during these interactions, but relevant observations will be logged which can later inform the assessor's interpretation of the submitted work.
		
		For future offerings, the more direct response is to increase the duration of practical sessions during demonstration weeks so that each team can be allocated a longer assessment window. A feasible target is approximately 15 minutes per team, comprising an eight-minute demonstration followed by around seven minutes of questioning. This would provide substantially greater opportunity to sample individual understanding while retaining the team demonstration format. More generally, where individual oral assurance is important, sufficient contact time must be treated as part of the assessment design rather than as an implementation detail.
	
	\subsection*{L4: Automation Scales Collection More Readily Than Interpretation}\label{sec:l4}
		Automation was essential to operating the redesign at cohort scale. Repository provisioning, milestone checking, autograding, and repository analysis reduced the manual effort required to create workspaces, collect evidence, and locate potentially relevant process information across hundreds of repositories. Without this infrastructure, repeated individual evidence collection would have been difficult to sustain, if not impossible.
		
		However, the first deployment also showed that automation addresses only part of the scaling problem. Determining whether a pull request represents meaningful review, whether a reflection is consistent with contribution traces, whether verification evidence is adequate, or whether contradictory evidence warrants escalation still requires human judgement. The same applies to formative feedback, oral questioning, scope review, and interpretation of unusual cross-channel patterns.
		
		The practical lesson is to distinguish \emph{collection scalability} from \emph{interpretation scalability}. The deployment demonstrated that substantial parts of evidence collection and aggregation could be automated; however, the architecture remains expensive to operate because interpretation, cross-checking, and escalation are attention-intensive activities. Replication therefore requires explicit workload allocation for these interpretive stages, not only technical infrastructure for collecting evidence.

\section{Design Implications and Transferability}\label{sec:transfer}
	The experience suggests a reframing of a common assessment-design question. Rather than asking ``\emph{How can this assessment be made AI-proof?}'', educators can ask ``\emph{What evidence would make the intended competence observable when AI assistance is available?}''. The answer depends on the competence being assessed. Authentic AI-assisted project work may provide strong evidence of integration, collaboration, verification, and engineering decision-making, while other outcomes may require more controlled individual evidence such as supervised practical work, oral questioning, or written examination.
	
	The architecture is therefore not an argument for universally permitting AI, nor does it prescribe a particular technology stack. Table \ref{tab:generalisability} illustrates how the architecture could be instantiated using lower-resource alternatives to the particular tooling used in this course. The essential requirement is not GitHub, Gradescope, or a custom AI-use declaration tool, but complementary evidence with different limitations and identifiable points of human responsibility.
	
	\begin{table*}[t]
	\centering
	\caption{Examples of how the architecture could be instantiated with different resources. These alternatives are design implications, not evaluated variants.}
	\label{tab:generalisability}
	\begin{tabular}{p{3.2cm} p{5.3cm} p{6.6cm}}
	\hline
	\textbf{Principle} & \textbf{Implementation in this course} & \textbf{Possible lower-resource implementation} \\
	\hline
	Process visibility & Individual GitHub repos, forks, PRs, issues & Shared GitLab/LMS version history; sampled commit review; structured change logs \\
	Verification & Gradescope, CI, simulator tests, reproducibility threshold & Instructor test harness; peer testing; sampled live checks \\
	Explanation & Weekly stand-ups, demo/Q\&A, written rationale & Sampled oral checks; short screencast defence; structured code walkthrough \\
	Process/origin & Custom AI Journal plus commit declarations & Standardised AI-use form or repository template \\
	Reflection & Marked individual reflection linked to evidence & Short evidence-linked reflection questions \\
	Individual assurance & Quizzes, milestones, secure final exam & Supervised practical, spot checks, or targeted oral examination \\
	Feedback loop & Aggregate AI-trace analysis in Weeks 4 and 8, later agentic-development instruction, and a planned Week 13 reflection & Manual sampling of representative interactions and cohort debrief \\
	\hline
	\end{tabular}
	\end{table*}

\section{Scope and Limitations}\label{sec:limitations}

	\paragraph{	Context and transferability}
	We described a single offering of a postgraduate, software-intensive engineering course at one institution. The course benefits from a technically favourable setting, including access to Git-based workflows, automated assessment infrastructure, and substantial coordinator involvement. The exact implementation may therefore not transfer directly to larger cohorts, less software-intensive disciplines, or settings with more limited technical and staffing resources. We consequently frame transferability primarily at the level of design principles rather than specific tools or assessment mechanisms.
	\paragraph{Evidence and effectiveness}
	There is no comparison condition, and no student-level learning outcomes, marks, or longitudinal behavioural change are analysed in this paper as the course is still running. The deployment observations therefore support lessons about the design and operation of the assessment system, but not causal claims that the redesign improved learning, engineering competence, or assessment validity.
	\paragraph{Status of the architecture}
	The AEA six evidence channels and CCE logic constitute a design decomposition rather than a validated measurement model of engineering competence. Each channel has known limitations, and the presence of multiple channels does not guarantee independent or reliable evidence. In particular, AI-assisted students may produce correlated evidence across code, documentation, reflections, and prepared explanations. The architecture should therefore be understood as a validity-oriented design strategy whose evidential claims require further empirical evaluation.
	\paragraph{Operational and reporting boundaries}
	The first deployment required substantial human effort despite extensive automation, particularly for feedback, cross-channel interpretation, and targeted follow-up. Visible assessment mechanisms may also influence student behaviour: AI-use records, repository activity, and reflections can become compliance-oriented when students know that these traces are assessed. Finally, we report the course at the stated reporting point; activities not yet completed are identified as planned rather than treated as deployment evidence. Finally, student-level records are not analysed in this paper.

\section{Conclusion}\label{sec:conclusion}
	When GenAI makes substantial software artefact production delegable, assessment can no longer rely on artefact quality alone as evidence of the competence behind it. This paper reports one whole-course response: structuring AI-assisted engineering so that students retain decision authority at planning, verification, review, and integration boundaries, while competence is inferred from complementary evidence distributed across the engineering and assessment process.
	
	The experience suggests that the practical unit of redesign is not the isolated ``AI-resistant'' task but the assessment system as a whole. Evidence must be collected across the engineering process, interpreted in relation to complementary sources, and followed up when important gaps or inconsistencies remain. The deployment also showed that the strength of such an architecture depends not only on the number of evidence channels, but on whether they provide genuinely different evidence, whether their intended role is realised under operational constraints, and whether sufficient human attention is available for interpretation.
	
	The architecture should therefore be understood as a design pattern and validity-oriented evidence strategy rather than a validated measurement model. Its central proposition is: when artefact authorship becomes less informative, educators need a more explicit argument about what evidence supports claims of engineering competence and where human responsibility remains visible.

\section*{Ethics} 
	This study was approved by the Human Research Ethics Committee of the University of Sydney (Approval No. HREC 2025/HE001132-substudy 022 v1b). This paper reports the design and operational experience of the course intervention and does not analyse or report student-level research data.

\section*{Data Availability}
	This experience report describes the design and operation of a course-level teaching and assessment intervention and does not analyse or report a student-level research dataset. Course artefacts and student records cannot be made publicly available because they contain participant and assessment information subject to institutional and ethics requirements. Supporting course tooling is being released progressively. The Visual Studio Code extension developed to integrate the AI Journal with Git-based workflows is already publicly and freely available as \emph{AI Journal Commit Form} through the Visual Studio Marketplace; additional course automation and analysis tooling is intended for public release after the conclusion of the semester. No student data will be included in these releases.

\section*{Acknowledgements}
	This project was supported by an Education-Focused Academics SoTL Grant from the University of Sydney. 

\bibliographystyle{IEEEtran}
\bibliography{references}

@inproceedings{shakiba_making_2026,
    address = {Sydney NSW Australia},
    title = {Making the {Invisible} {Visible}: {A} {Framework} for {Reflective} {AI} use in {Software} {Engineering} {Education}},
    publisher = {Australasian Association for Engineering Education},
    author = {Shakiba, Ali and Chaffey, Thomas},
    month = dec,
    year = {2026},
}

@article{bearman2024evaluative,
	author = {Margaret Bearman and Joanna Tai and Phillip Dawson and David Boud and Rola Ajjawi},
	title = {Developing evaluative judgement for a time of generative artificial intelligence},
	journal = {Assessment \& Evaluation in Higher Education},
	volume = {49},
	number = {6},
	pages = {893-905},
	year = {2024},
	publisher = {Routledge},
	doi = {10.1080/02602938.2024.2335321},
}

@article{ebrahimzadeh2026coauthorship,
	title = {Coauthorship integrity: Reconceptualising assessment validity for the age of generative artificial intelligence},
	journal = {Computers and Education: Artificial Intelligence},
	volume = {10},
	pages = {100609},
	year = {2026},
	issn = {2666-920X},
	doi = {https://doi.org/10.1016/j.caeai.2026.100609},
	url = {https://www.sciencedirect.com/science/article/pii/S2666920X26000718},
	author = {Mohsen Ebrahimzadeh and Antonette Shibani and Simon Buckingham Shum},
}

@article{bassner2026less,
	title = {Less stress, better scores, same learning: The dissociation of performance and learning in {AI}-supported programming education},
	journal = {Computers and Education: Artificial Intelligence},
	volume = {10},
	pages = {100537},
	year = {2026},
	issn = {2666-920X},
	doi = {https://doi.org/10.1016/j.caeai.2025.100537},
	author = {Patrick Bassner and Ben Lenk-Ostendorf and Ramona Beinstingel and Tobias Wasner and Stephan Krusche},
}

@article{biggs1996constructive,
	author  = {John Biggs},
	title   = {Enhancing Teaching through Constructive Alignment},
	journal = {Higher Education},
	volume  = {32},
	number  = {3},
	pages   = {347-364},
	year    = {1996},
	doi     = {10.1007/BF00138871}
}

@article{gulikers2004authentic,
  title={A five-dimensional framework for authentic assessment},
author={Gulikers, Judith TM and Bastiaens, Theo J and Kirschner, Paul A},
journal={Educational technology research and development},
volume={52},
number={3},
pages={67-86},
year={2004},
publisher={Springer},
doi={10.1007/BF02504676}
}

@article{vandervleuten2010competence,
title = {The assessment of professional competence: building blocks for theory development},
journal = {Best Practice \& Research Clinical Obstetrics \& Gynaecology},
volume = {24},
number = {6},
pages = {703-719},
year = {2010},
note = {Training, Education and Assessment},
issn = {1521-6934},
doi = {https://doi.org/10.1016/j.bpobgyn.2010.04.001},
url = {https://www.sciencedirect.com/science/article/pii/S1521693410000519},
author = {C.P.M. {van der Vleuten} and L.W.T. Schuwirth and F. Scheele and E.W. Driessen and B. Hodges},
}

@inproceedings{buffardi2020gitlogs,
author = {Buffardi, Kevin},
title = {Assessing Individual Contributions to Software Engineering Projects with {Git} Logs and User Stories},
year = {2020},
isbn = {9781450367936},
publisher = {Association for Computing Machinery},
address = {New York, NY, USA},
url = {https://doi.org/10.1145/3328778.3366948},
doi = {10.1145/3328778.3366948},
booktitle = {Proceedings of the 51st {ACM} Technical Symposium on Computer Science Education},
pages = {650-656},
numpages = {7},
location = {Portland, OR, USA},
series = {SIGCSE '20}
}

@article{fernandezgauna2023devops,
title={Automatic feedback and assessment of team-coding assignments in a {DevOps} context}, 
volume={20}, 
ISSN={2365-9440}, 
url={http://dx.doi.org/10.1186/s41239-023-00386-6}, 
DOI={10.1186/s41239-023-00386-6}, 
number={1}, 
journal={International Journal of Educational Technology in Higher Education}, 
publisher={Springer Science and Business Media LLC}, 
author={Fernandez-Gauna, Borja and Rojo, Naiara and Graña, Manuel},
 year={2023}, 
month=Mar }

@inproceedings{fein2026aidriven,
	author = {Fein, Benedikt and Fraser, Gordon and Herbold, Steffen},
	title = {{AI}-Driven Software Development: A New Course Concept and Assessment Model for the Era of Large Language Models},
	year = {2026},
	isbn = {9798400724237},
	publisher = {Association for Computing Machinery},
	address = {New York, NY, USA},
	url = {https://doi.org/10.1145/3786580.3786962},
	doi = {10.1145/3786580.3786962},
	booktitle = {Proceedings of the {IEEE/ACM} 48th International Conference on Software Engineering},
	pages = {386-396},
	numpages = {11},
	location = {Rio de Janeiro, Brazil},
	series = {ICSE-SEET '26}
	}

@article{perkins2024aias, 
	place={Launceston, Australia}, 
	title={The Artificial Intelligence Assessment Scale ({AIAS}): A Framework for Ethical Integration of Generative {AI} in Educational Assessment}, 
	volume={21}, 
	DOI={10.53761/q3azde36}, 
	number={06}, 
	journal={Journal of University Teaching and Learning Practice}, 
	author={Perkins, Mike and Furze, Leon and Roe, Jasper and MacVaugh, Jason}, 
	year={2024}, 
	month={Apr} 
}

@book{two_lane_2023,
	title = "Assessment Reform for The Age of Artificial Intelligence",
	author = "Jason Lodge and Sarah Howard and Bearman, Margaret Lynn and Phillip Dawson and Shirley Agostinho and Buckingham Shum, Simon J. and Chris Deneen and Cath Ellis and Tim Fawns and Helen Gniel and Rowena Harper and Michael Henderson and Liu, Danny Yen Ting and Lina Markauskaite and Jan McLean and Carlo Perrotta and Lambert Schuwirth and Christine Slade and Associates",
	year = "2023",
	month = nov,
	language = "English",
	publisher = "Tertiary Education Quality and Standards Agency",
}

@article{mislevy2003structure,
	author = {Robert J. Mislevy and Linda S. Steinberg and Russell G. Almond},
	title = {Focus Article: On the Structure of Educational Assessments},
	journal = {Measurement: Interdisciplinary Research and Perspectives},
	volume = {1},
	number = {1},
	pages = {3-62},
	year = {2003},
	publisher = {Routledge},
	doi = {10.1207/S15366359MEA0101\_02},
}

@article{brattli2026validity,
  title   = {Assessment Validity in the Age of Generative AI: A Natural Experiment},
  volume={13}, ISSN={2227-9709}, url={http://dx.doi.org/10.3390/informatics13040056}, DOI={10.3390/informatics13040056}, number={4}, journal={Informatics}, publisher={MDPI AG}, author={Brattli, Havar and Utne, Alexander and Lynch, Matthew}, year={2026}, month=Apr, pages={56} 
}

@article{kuhail2024replaced,
  title={"Will I be replaced?" {A}ssessing {ChatGPT}'s effect on software development and programmer perceptions of {AI} tools}, volume={235}, ISSN={0167-6423}, url={http://dx.doi.org/10.1016/j.scico.2024.103111}, DOI={10.1016/j.scico.2024.103111}, journal={Science of Computer Programming}, publisher={Elsevier BV}, author={Kuhail, Mohammad Amin and Mathew, Sujith Samuel and Khalil, Ashraf and Berengueres, Jose and Shah, Syed Jawad Hussain}, year={2024}, month=jul, pages={103111} 
}

@article{weberwulff2023testing,
  title   = {Testing of Detection Tools for {AI}-Generated Text},
  volume={19}, ISSN={1833-2595}, url={http://dx.doi.org/10.1007/s40979-023-00146-z}, DOI={10.1007/s40979-023-00146-z}, number={1}, journal={International Journal for Educational Integrity}, publisher={Springer Science and Business Media LLC}, author={Weber-Wulff, Debora and Anohina-Naumeca, Alla and Bjelobaba, Sonja and Foltýnek, Tomáš and Guerrero-Dib, Jean and Popoola, Olumide and Šigut, Petr and Waddington, Lorna}, year={2023}, month=Dec 
}

@article{sun2026trusting,
  title   = {Trusting AI to Detect AI? A Systematic Evaluation of the Reliability and Robustness of Current {AIGC} Detection Tools for Student Academic Work},
  volume={249}, ISSN={0360-1315}, url={http://dx.doi.org/10.1016/j.compedu.2026.105616}, DOI={10.1016/j.compedu.2026.105616}, journal={Computers \& Education}, publisher={Elsevier BV}, author={Sun, Yicheng and Liao, Yihan and Ma, Xiaoxue}, year={2026}, month=Aug, pages={105616}
}

\end{document}